\documentclass[journal]{IEEEtran}
\ifCLASSINFOpdf
\else
\fi
\usepackage{graphicx} 
\usepackage{float} 
\usepackage{subfigure} 
\usepackage{diagbox}
\usepackage{algorithm}
\usepackage{algorithmic}
\usepackage{multirow}
\usepackage{amsmath}
\usepackage{caption}
\usepackage{cite}
\usepackage{color}
\usepackage{url,array}
\usepackage{hyperref}
\usepackage{booktabs}

\begin{document}
%
\title{Timbre-reserved Adversarial Attack in Speaker Identification}
%
%
%

\author{Qing~Wang,
        Jixun~Yao,
        Li~Zhang,
        Pengcheng~Guo,
        and~Lei~Xie,~\IEEEmembership{Senior member,~IEEE} \thanks{Manuscript received December 27, 2022; revised May 31, 2023; accepted August 10, 2023. \textit{(Corresponding author: Lei Xie.)}} \thanks{Qing Wang, Jixun Yao, Li Zhang, Pengcheng Guo, and Lei Xie are with the Audio, Speech and Langauge Processing Group (ASLP@NPU), School of Computer Science, Northwestern Polytechnical University, Xi’an 710129, China (e-mail: qingwang@nwpu-aslp.org; yaojx@mail.nwpu.edu.cn; alivacheung@gmail.com; guopengcheng1220@gmail.com; lxie@nwpu.edu.cn).}}

%
%

\markboth{Journal of \LaTeX\ Class Files,~Vol.~14, No.~8, August~2015}%
{Shell \MakeLowercase{\textit{et al.}}: Bare Demo of IEEEtran.cls for IEEE Journals}
%



\maketitle

\begin{abstract}
 As a type of biometric identification, a speaker identification (SID) system is confronted with various kinds of attacks. The spoofing attacks typically imitate the timbre of the target speakers, while the adversarial attacks confuse the SID system by adding a well-designed adversarial perturbation to an arbitrary speech. Although the spoofing attack copies a similar timbre as the victim, it does not exploit the vulnerability of the SID model and may not make the SID system give the attacker's desired decision. As for the adversarial attack, despite the SID system can be led to a designated decision, it cannot meet the specified text or speaker timbre requirements for the specific attack scenarios. In this study, to make the attack in SID not only leverage the vulnerability of the SID model but also reserve the timbre of the target speaker, we propose a timbre-reserved adversarial attack in the speaker identification. We generate the timbre-reserved adversarial audios by adding an adversarial constraint during the different training stages of the voice conversion (VC) model. Specifically, the adversarial constraint is using the target speaker label to optimize the adversarial perturbation added to the VC model representations and is implemented by a speaker classifier joining in the VC model training. The adversarial constraint can help to control the VC model to generate the speaker-wised audio. Eventually, the inference of the VC model is the ideal adversarial fake audio, which is timbre-reserved and can fool the SID system. Experimental results on the Audio deepfake detection (ADD) challenge dataset indicate that our proposed method improves the attack success rate significantly compare with the vanilla VC model without additionally introducing an adversarial noise to the attack speech. Objective and subjective evaluations illustrate that the quality of fake audio generated by our proposed method is better than directly adding adversarial perturbation to the VC-generated audio. Furthermore, the analysis shows that our generated adversarial fake audios also meet the specified text and target speaker timbre-reserved requirements of the attacker.

\end{abstract}

\begin{IEEEkeywords}
Speaker identification, adversarial attack, timbre-reserved, voice conversion.
\end{IEEEkeywords}

%
\IEEEpeerreviewmaketitle

\section{Introduction}
Speaker identification (SID)~\cite{hansen2015speaker,reynolds1995robust} is the process of automatically inferring the identity of a speaker from a spoken utterance. As one of the most prominent biometric authentication methods, it is critical to ensure the robustness of the SID system.
A variety of attacks attempt to challenge the robustness of the SID system. 
For instance, spoofing attack~\cite{wu2012detecting, wu2015spoofing, wu2017asvspoof} commonly includes impersonation, replay, voice conversion, and speech synthesis. 
The ASVspoof challenge~\cite{wu2015asvspoof, kinnunen2017asvspoof, todisco2019asvspoof} is dedicated to addressing this problem, and it is organized to contribute to the development of countermeasures to protect speaker recognition from the threat of spoofing attacks. 
Depending upon the scenarios of the spoof samples attacking the SID system, the spoofing attack can be broadly classified into two categories, which are physical access (PA) attacks and logical access (LA) attacks~\cite{todisco2019asvspoof}. 
In the PA attacks, the samples are applied as input to the SID system through the sensor, which directly attacks the SID system. 
The LA attacks are the direct injection into the SID system, and the most common approaches are speech synthesis and voice conversion. Both of them aim to generate a speech based on the voice of the target speaker. 
However, a spoofing attack normally depends on the quality and size of the dataset and cannot take advantage of SID model's vulnerability, so that the spoofing attack is difficult to accurately be classified to the target speaker by the SID system since the spoofing attacks do not consider the downstream SID task.

Moreover, the adversarial attack~\cite{goodfellow2014explaining, carlini2017towards, szegedy2013intriguing, kurakin2016adversarial} is considered as another type of attack, which is emerged in the last few years. An adversarial attack is a malicious attempt that exploits the vulnerability of the network itself and tries to perturb the original sample into a new sample. The new sample, which is also known as an adversarial example, can be misclassified by the network. Adversarial attacks also can impact speech processing tasks~\cite{carlini2018audio, schonherr2018adversarial, alzantot2018did, sun2019adversarial, qin2019imperceptible, abdullah2021sok}. 
Since recent work has explored different deep neural network (DNN) architectures~\cite{snyder2018x, wan2018generalized, desplanques2020ecapa} to produce compact speaker embeddings, the SID system is also vulnerable to adversarial examples.
In recent years, many researchers have successfully conducted adversarial attacks on SID systems~\cite{kreuk2018fooling, wang2019adversarial, abdullah2019practical, li2020universal, xie2020real, li2020practical, li2020adversarial, chen2021real, jati2021adversarial}. 
However, the perturbation is added to an arbitrary speech to achieve the targeted adversarial attack, which is hard to meet the requirement of speaker similarity and intelligibility and can easily be detected by humans.
What's more, although the perturbation is usually designed too small to be heard by human beings through multiple approaches~\cite{wang2020inaudible, zhang2022imperceptible}, the adversarial perturbation is not strictly inaudible.

Inspired by these two kinds of attacks above, qualified fake audio for attacking the SID model should have the ability to deceive both machines and humans at the same time. 
To fool the machine, the downstream SID task needs to be considered so that the fake audio has a distinctive speaker attribute, which can make the SID model make the designated decision.
From a human perception perspective, when the timbre or the text of the fake audio is far different from the target speaker's real audio, this kind of fake audio is easy to be detected. Therefore, the timbre and text information also are concerns when we conduct attacks on the SID model. 
Moreover, the quality of the fake audio is also crucial in the attack.

In this study, when we conduct an adversarial attack on the speaker identification model, we aim to take the target speaker's timbre and text information into consideration. To this end, we propose a timbre-reserved adversarial attack in the SID system, which is to make the attack in SID not only exploit the vulnerability of the SID model but also reserve the timbre of the target speaker, as well as customize the text of the fake speech. 
In particular, a speaker classifier is jointly trained with the VC model to determine whether the representations belong to the target speaker. If it is not classified into the target speaker class, an adversarial perturbation is added to the VC model representation, which constrains the representations to be classified to the target speaker. Then the perturbed representation proceeds with optimizing the VC model.
Consequently, the VC model can generate fake audio with distinctive target speaker information.
We adopt various levels of VC model representation during the VC model training.  
In non-autoregressive based VC model training, the adversarial constraint is added to the Mel-spectrogram and latent representation. While in the end-to-end VC model training, we add the adversarial constraint to the reconstructed waveform.
After that, by adding adversarial constraints to optimize various voice conversion frameworks, we can obtain the timbre-reserved and speaker-wised adversarial fake audios for attacking the speaker identification system. 

We evaluate our proposed methods on the Audio deepfake detection (ADD) challenge dataset~\cite{yi2022add}. With the given text and speakerID, the fake audios generated by our proposed method improve the attack success rate significantly compared with the vanilla VC model. Since the proposed methods do not introduce extra adversarial noise to the attack speech, the objective and subjective evaluations also illustrate that the quality of fake audio generated by our proposed method is better than directly adding adversarial perturbation to the VC generated audio.
Furthermore, we also analyze the speaker similarity and intelligibility of the fake audio, which both meet the requirements.

We summarize our main contributions as follows.
\begin{itemize}
\item To our best knowledge, we are the first to propose adding adversarial perturbation generation into voice conversion framework training to achieve timbre-reserved adversarial attacks for the speaker identification system.
\item We explore adding adversarial constraints to different levels of representation of the VC training, which are the VC predicted Mel-spectrogram, latent representation, and reconstructed waveform, respectively.
\item The timbre-reserved fake audio not only preserves the timbre of the target speaker but also can effectively targeted attack the SID system, and the text of the fake audio can be customized as well.
\end{itemize}
The rest of the paper is organized as follows.
In Section \uppercase\expandafter{\romannumeral2}, related works are introduced. 
In Section \uppercase\expandafter{\romannumeral3}, we detail the proposed timbre-reserved adversarial attack in the SID system.
Datasets and experimental setup are described in Section \uppercase\expandafter{\romannumeral4}.
Section \uppercase\expandafter{\romannumeral5} presents the experimental results and analysis.
We conclude in Section \uppercase\expandafter{\romannumeral6}.

\section{Related works}
\subsection{Spoofing attacks in speaker identification}
In recent years, there are various kinds of attacks on speaker identification. One of the most severe threats is a spoofing attack. The spoofed speech samples can be obtained through impersonation, replay, voice conversion, or speech synthesis. Especially, speech synthesis and VC are also the main concerns as typical types of deep fake in SID and are the logical access (LA) tasks in spoofing attacks as well.

Speech synthesis, also known as text-to-speech (TTS), takes arbitrary text as input and generates speech as output~\cite{wang2017tacotron, ren2019fastspeech}. With the development of the TTS techniques, synthesized speech is more indistinguishable from human speech. 
In~\cite{chen2010speaker, shchemelinin2014vulnerability, wu2015asvspoof, wu2017asvspoof}, TTS-generated fake audios were used as the spoofing attack in speaker recognition systems.
However, the attack on the speaker recognition system using the TTS-generated fake audio is not very effective, since the downstream SID task is not taken into consideration. In~\cite{cai2018attacking}, Cai \textit{et al.} proved that samples generated with SampleRNN~\cite{mehri2016samplernn} and WaveNet~\cite{oord2016wavenet} were unable to fool deep learning based speaker recognition system. Wenger \textit{et al.}~\cite{wenger2021hello} generated synthetic speech using publicly available systems that can already fool both humans and several popular software systems, but the performance of the attack success rate is not that high enough.
In addition, the quality and size of the training set for the TTS model are typically insufficient for genuine attacking a SID model, as a result, it is difficult to attack the SID system with speech synthesis itself.

Voice conversion (VC) is a process that converts or transforms the voice of the original speaker into the target speaker while keeping the linguistic content~\cite{sisman2020overview}. In this study, we mainly focus on the generation of fake audio by VC. 
With the development of VC, different kinds of VC models have been explored, and they are typically divided into three types: autoregressive based VC model~\cite{sun2015voice,sun2016phonetic}, non-autoregressive based VC model~\cite{hayashi2021non,liu2020non,hayashi2022investigation,kameoka2018stargan,kaneko2018cyclegan}, and end-to-end VC model~\cite{lin2021fragmentvc,nguyen2022nvc,kim2021conditional}.

The first type of VC model is autoregressive based. 
The typical autoregressive based VC model~\cite{sun2015voice,sun2016phonetic} is an improved version of Tacotron 2~\cite{shen2018natural}, which leverages an encoder-decoder framework and takes Mel-spectrograms or the bottleneck features of automatic speech recognition (ASR) model as inputs. 
However, the autoregressive VC model predicts Mel-spectrogram frame by frame and can not train in a parallel way, which is unsuitable as the baseline model of adversarial attack. As a result, we do not use it as our VC baseline model.

Secondly, various non-autoregressive VC models capable of synthesizing speech in a non-autoregressive manner have been proposed to speed up the training and inference process enormously. The non-autoregressive models generate parallel sequences, solving word skipping and repetition problems caused by incorrect attention alignment. The typical non-autoregressive VC model is based on FastSpeech 2~\cite{fs2}, and consists of multiple feed-forward transformer (FFT) blocks.
Mel-spectrograms generated using a non-autoregressive method do not rely on the previous frames and are suitable for adding adversarial constraints.
Therefore, the FastSpeech-based VC model is used as our first VC baseline model for generating fake audio.

The third type is fully end-to-end (E2E) VC models, which can generate speech waveform from waveform or bottleneck features directly. 
Different from the previous two kinds of VC models, the E2E VC model requires less human annotation and feature development, while the joint optimization of the E2E VC model can avoid the distribution mismatch between the acoustic model and the vocoder.
Moreover, the E2E VC model can also reduce training and deployment costs. 
In this study, to comprehensively investigate the effect of adversarial constraints for different levels of representations of the VC models, we adopt the HifiGAN-based E2E VC model~\cite{kashkin2022hifi} as our second baseline model. 

Different from the TTS-based fake audio generation, the VC model does not have high requirements on the quality and size of the training data. However, neither the TTS nor the VC model takes the downstream SID task into consideration, and the performance of attacking the SID model by spoofed audios is not as expected.
To better attack the SID model, besides preserving the timbre of the target speaker, the attack needs to leverage the vulnerability of the SID network as well.

\subsection{Adversarial attack in speaker identification}
The adversarial attack is typically divided into two categories according to whether the attackers expect the model is misleading to a specific output, which are targeted and non-targeted attacks. We mainly focus on targeted attacks in speaker identification tasks in this study.
Das \textit{et al.}~\cite{das2020attacker} gave an overview of various types of attack on speaker verification focusing on potential threats of adversarial attacks and spoofing countermeasures from the attacker's perspective.
In~\cite{li2020universal}, Li \textit{et al.} explored the existence of the universal adversarial perturbations (UAPs) in speaker recognition systems and proposed to generate UAPs by learning the mapping from the low-dimensional normal distribution to the UAP subspace via a generative model.
Xie \textit{et al.}~\cite{xie2020real} proposed the real-time, universal, and robust adversarial attack against DNN-based speaker recognition system by adding audio-agnostic universal perturbations. 
Li \textit{et al.} ~\cite{li2020practical} launched a practical and systematic adversarial attack against speaker recognition systems and integrated the estimated room impulse response into the adversarial example training for over-the-air attack. 
In our previous work~\cite{wang2020inaudible}, to targeted attack the speaker recognition system, we generated inaudible adversarial perturbations based on psychoacoustic principle of frequency masking.
Zhang \textit{et al.}~\cite{zhang2022imperceptible} performed black-box waveform-level targeted adversarial attacks against speaker recognition systems by generating imperceptible adversarial perturbations based on auditory masking. 
In~\cite{chen2021real}, Chen \textit{et al.} proposed FAKEBOB to craft adversarial examples and conducted a comprehensive and systematic study of the adversarial attacks on speaker recognition systems to understand their security weakness in the practical black-box setting. 

However, to achieve a targeted adversarial attack, all the studies mentioned above add adversarial perturbation to an arbitrary speech. From a human perception perspective, these adversarial audio sounds completely different from the target speaker's timbre. When the attack needs to meet the requirement of text and speaker similarity, these adversarial attack is easy to be detected. Furthermore, even though these methods introduce small perturbations as inaudible as possible to the fake audio, the perturbations are usually directly added to the waveform and it can still be perceived by humans.

\section{Methodology}\label{sec:method}

In this section, we introduce how we generate timbre-reserved fake audio.
We add adversarial constraints in different levels of representations of voice conversion model training: VC predicted Mel-spectrogram, latent representation, and reconstructed waveform.

\subsection{Adversarial constraint}
To make the fake audios generated by the VC model have distinctive speaker attributes, we add an adversarial constraint process into VC model training, which can adversely restrict the VC model representations and lead to the speaker classifier giving the target prediction. 
In the adversarial constraint process, suppose that a well-trained speaker identification model is $f(\cdot)$, and the Mel-spectrogram $M$ is the input of the SID model, which is the VC model predicted representation. And its corresponding speaker label predicted by the SID model $f(\cdot)$ is $y$, while its VC target speaker label is $y'$. When $y=y'$, the Mel-spectrogram $M$ generated from the VC model has the right speaker attribute. In contrast, when $y \neq y'$, the adversarial constraint process is used to modify $M$ generated from VC model, and the adversarial constraint $\delta$ can be defined as follows:
\begin{equation}\label{commo_method}
    \begin{aligned}
        & \min L_{CE}(f(M+\delta), y'), \\
        & \text { s.t. } \quad \lVert \delta \rVert < \epsilon,        
    \end{aligned}
\end{equation}
where $L_{CE}(\cdot)$ is the loss function to make adversarial examples lead the SID model predicting the specific target speaker label, and the hyperparameter $\epsilon$ is used to control the maximum perturbation generated. The $\delta$ is initialized to a zero vector and $\epsilon$ is gradually reduced from a large value. For each iteration, $\delta$ is updated as follows:
\begin{equation}\label{lr}
    \delta \leftarrow \text{clip}_{\epsilon}(\delta - lr \cdot \text{sign}(\nabla_{\delta}L_{CE}(f(M+\delta), y'))),
\end{equation}
where $lr$ is the learning rate, $\nabla_{\delta}L_{CE}$ is the gradient of SID model with respect to $\delta$.
When we attempt to perturb the original VC model, this adversarial constraint is added in various VC representations during the VC model training to reserve the distinctive target speaker information.

\subsection{Adversarial constraint on Mel-spectrogram}

\begin{figure*}[ht]
  \centering
  \includegraphics[width=16.5cm]{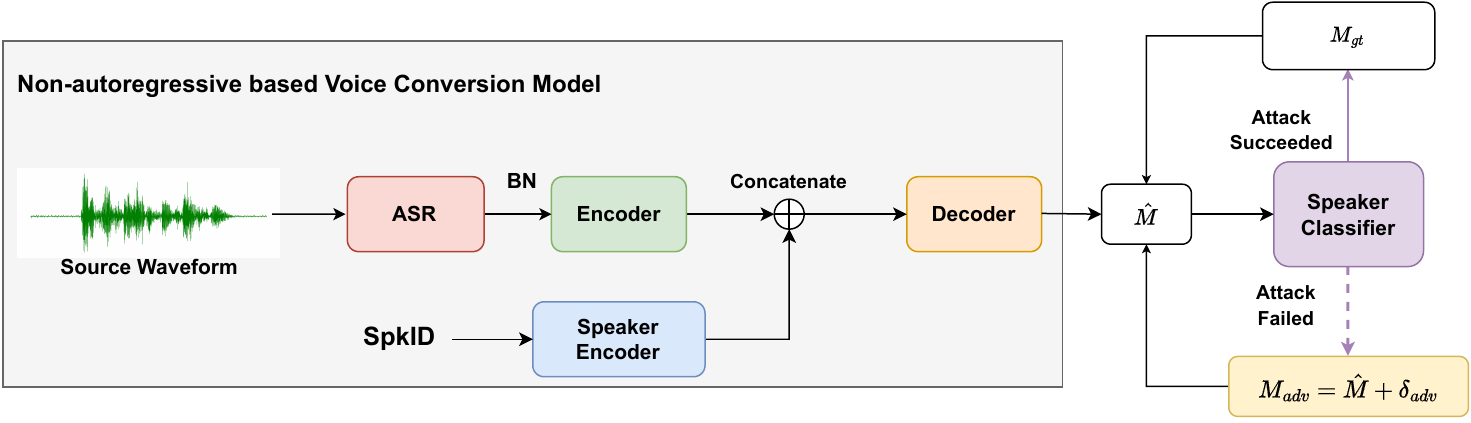}
  \caption{An overview of adversarial constraint on Mel-spectrogram. The gray box is the non-autoregressive based voice conversion model, whose inputs are source wave and speaker ID. The predicted Mel-spectrogram is then classified by the speaker classifier, and it is determined if it is the target speaker to decide whether to add the adversarial constraint.}
  \label{fig:vcadv_1}
\end{figure*}

Figure~\ref{fig:vcadv_1} is an overview of our first strategy of adversarial constraint on Mel-spectrogram, which consists of a non-autoregressive based voice conversion model and an attack constraint process. The VC model is trained beforehand and then jointly trained with the attack constraint process.
In order to make the VC model predict the Mel-spectrogram with a distinctive target speaker attribute, we conduct our first strategy in the following steps: first, the voice conversion model converts the source speaker's identity to the target speaker $y'$; then, the adversarial constraint is added to optimize the predicted Mel-spectrogram when the prediction of speaker classifier is not as same as the target speaker.

For VC model training, a pre-trained ASR model is first used to extract speaker-independent linguistic information from the source waveform.
We employ the bottleneck feature extracted from the ASR final encoder layer as linguistic information. 
The acoustic model predicts the Mel-spectrogram by a 6-layer conformer encoder-decoder structure similar to FastSpeech 2~\cite{fs2}. The speaker embedding from the speaker encoder is fed to the decoder as conditions for target voice generation.
We adopt the L1 loss as reconstruction loss to optimize the VC model and is defined as:
\begin{equation}\label{vc_loss}
    \mathcal{L}_{\texttt{rec}}=||M_{\texttt{gt}}-\hat{M}||_{1},
\end{equation}
where $M_{\texttt{gt}}$ and $\hat{M}$ represent the ground truth Mel-spectrogram of source speech and the VC predicted Mel-spectrogram, respectively.

As for joint training with attack constraint, we use the Mel-spectrogram predicted by the VC model to attack the speaker classifier, which is a well-trained speaker identification model sharing the same speaker set as the VC training data. 
If the attack fails, as shown in the right part of Figure~\ref{fig:vcadv_1}, we add a tiny adversarial perturbation to the predicted Mel-spectrogram to generate the adversarial Mel-spectrogram $M_{\texttt{adv}}$ with an adversarial constraint which can be defined as: 
\begin{equation}
\begin{aligned}
& M_{\texttt{adv}}=\hat{M}+\delta_{\texttt{adv}} \\
&\text { s.t. } \quad\|\delta_{\texttt{adv}}\|<\epsilon ,
\end{aligned}
\end{equation}
here, $\delta_{\texttt{adv}}$ represents the tiny adversarial perturbation and $\epsilon$ is used to control the maximum adversarial perturbation generated. The tiny adversarial perturbation can be optimized by: 
\begin{equation}
\begin{aligned}
\min \mathcal{L}_{\texttt{CE}}\left(f(M_{\texttt{adv}}), y^{\prime}\right) ,
\end{aligned}
\end{equation}
where $\mathcal{L}_{\texttt{CE}}$ aims to make the $M_{\texttt{adv}}$ fool the well-trained speaker identification system into predicting a specified target label. 
Therefore, the joint training with adversarial constraint can be optimized by the following loss function: 
\begin{equation}\label{tot_loss}
    \mathcal{L}_{\texttt{adv}}=\left\{  
        \begin{array}{lr}  
             ||M_{\texttt{gt}}-\hat{M}||_{1}, \quad \text{if succeeded}, &  \\  
             ||M_{\texttt{adv}}-\hat{M}||_{1}, \quad \text{if failed}. &    
        \end{array}  \right.
\end{equation}
A well-trained speaker classifier $f(\cdot)$ is added behind the VC model, the Mel-spectrogram predicted from the VC model $\hat{M}$ is used as the input of $f(\cdot)$, and the system determines whether the prediction of the speaker classifier $f(\hat{M})$ is the target speaker $y'$ or not.
When $f(\hat{M}) \neq y'$, in other words, the attack failed, we add an adversarial perturbation to the Mel-spectrogram as the adversarial constraint.
In order to force the predicted Mel-spectrogram $\hat{M}$ of the VC model can be classified to the target speaker, we expect to minimize the L1 loss between the predicted Mel-spectrogram $\hat{M}$ and the Mel-spectrogram with the adversarial perturbation $M_{\texttt{adv}}$ so that the VC model can fool the well-trained speaker recognition system.
The adversarial perturbation $\delta_{\texttt{adv}}$ is optimized by Equation~\ref{lr} 
until the predicted label $f(\hat{M})$ is the target label. 
In contrast, when the speaker classifier gives a prediction of the target speaker label, which means the attack succeeds, the VC model is optimized only using the original reconstruction loss in Equation~\ref{vc_loss}.

After VC model joint training with the attack constraint process, the generation of deep fake audio is based on the HifiGAN vocoder~\cite{kong2020hifi}, whose input is the Mel-spectrogram predicted by the adversarial constrained VC model.

\subsection{Adversarial constraint on latent representation}

\begin{figure*}[t]
  \centering
  \includegraphics[width=17.5cm]{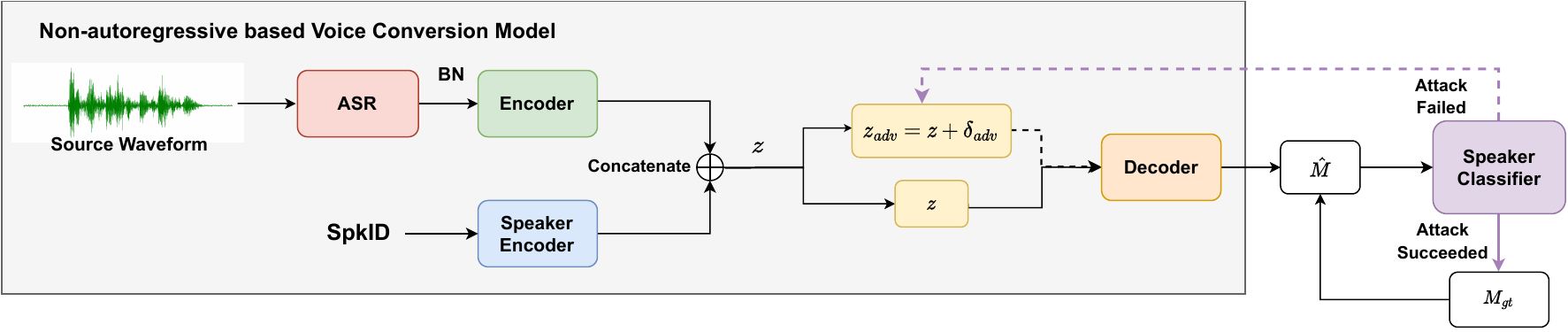}
  \caption{Adversarial constraint on latent representation. The gray box is the non-autoregressive based voice conversion model. The latent representation is the concatenation of the high-level linguistic representation and speaker embedding. }
  \label{fig:overview2}
\end{figure*}

As shown in Figure~\ref{fig:overview2}, we add the adversarial constraint to the latent representation of the VC model in the second strategy.
The encoder takes the bottleneck (BN) feature as the input and outputs the high-level linguistic representation, while the  speaker encoder is adopted to generate speaker embedding.
The latent representation $z$ is the concatenation of the high-level linguistic representation and speaker embedding. The adversarial constraint on latent representation $z_{\texttt{adv}}$ is operated by adding the adversarial perturbation, which can be formulated as follow:
\begin{equation}
    z_{\texttt{adv}}=z+\delta_{\texttt{adv}}.
\end{equation}
The following optimization of tiny adversarial perturbation and joint training is as same as the first strategy: 
\begin{equation}
\begin{aligned}
&\min L_{C E}\left(f(Dec(z_{\texttt{adv}})), y^{\prime}\right), \\
\end{aligned}
\end{equation}
where $Dec(\cdot)$ represents the FFT-based decoder and $Dec(z_{\texttt{adv}})$ is the Mel-spectrogram predicted by the FFT-based decoder from the adversarial perturbed latent representation. Next, the second strategy of adversarial constraint on latent representation can be optimized by: 
\begin{equation}
    \mathcal{L}_{\texttt{adv}}=\left\{  
        \begin{array}{lr}  
             ||M_{\texttt{gt}}-\hat{M}||_{1}, \quad \text{if succeeded}, &  \\  
             ||Dec(z_{\texttt{adv}})-\hat{M}||_{1}, \quad \text{if failed}. &    
        \end{array}  \right.
\end{equation}
The adversarial perturbation is added to the latent representation when $f(\hat{M}) \neq y'$. And we minimize the difference of predicted Mel-spectrogram $\hat{M}$ and the Mel-spectrogram $Dec(z_{\texttt{adv}})$ decoded from the perturbed latent representation.
The adversarial perturbation $\delta_{\texttt{adv}}$ is optimized by Equation~\ref{lr} until the predicted label $f(\hat{M})$ is the target label $y'$. 
When the attack succeeds, the VC model is optimized as normal in Equation~\ref{vc_loss}.

\subsection{Adversarial constraint on reconstructed waveform}

\begin{figure}[t]
  \centering
  \includegraphics[width=5.5cm]{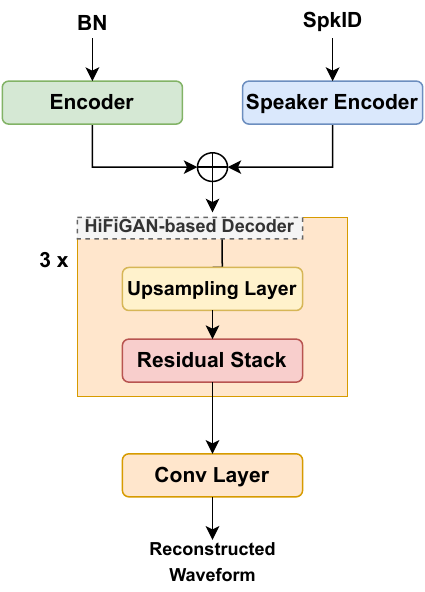}
  \caption{The architecture of end-to-end voice conversion model. }
  \label{fig:E2emodel}
\end{figure}

\begin{figure*}[ht]
  \centering
  \includegraphics[width=17.5cm]{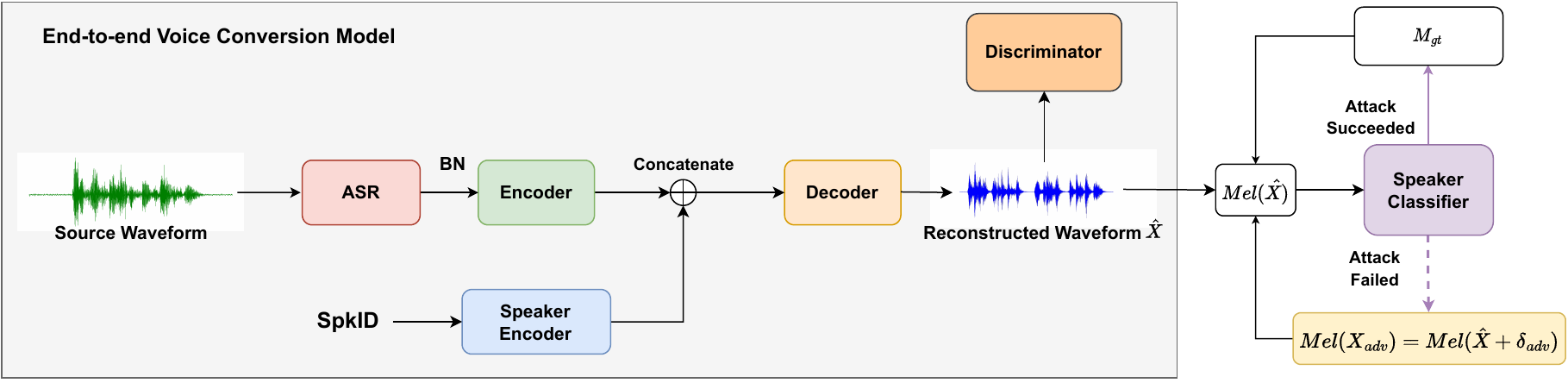}
  \caption{Adversarial constraint on the reconstructed waveform. The gray box is the end-to-end based voice conversion model. The adversarial perturbation is added to the reconstructed waveform if the reconstructed waveform is not classified as the target speaker by the speaker classifier. }
  \label{fig:overview3}
\end{figure*}
In addition to adding adversarial perturbation on Mel-spectrogram or latent representation, we also add adversarial perturbation on the reconstructed waveform. An end-to-end HiFiGAN-based voice conversion model replaces the VC model and the detailed architecture is illustrated in Figure~\ref{fig:E2emodel}, the VC module goes through single-stage training for efficient end-to-end learning. The VC model consists of a convolutional long short-term memory (CLSTM) based encoder, a speaker encoder, and a HiFiGAN-based decoder~\cite{kong2020hifi}, which aim at high-level linguistic representation encoding and waveform reconstruction, respectively. CLSTM consists of three stacks of convolution layers, the LeakyReLU activation function, and an LSTM layer. The speaker embedding from the lookup table is fed to the decoder as conditions for target voice generation. The architecture of the decoder generator follows the same configuration as HiFi-GAN~\cite{kong2020hifi}.

The training objective of the end-to-end VC HiFiGAN~\cite{kashkin2022hifi}, which consists of reconstruction loss $\mathcal{L}_{\texttt{rec}}$, feature matching loss $\mathcal{L}_{\texttt{fm}}$, and adversarial loss $\mathcal{L}_{\texttt{vc-adv}}$. 
As for reconstruction loss, we compute L1 loss between spectrograms of the source waveform $X_{\texttt{s}}$ and predicted waveform $\hat{X}_{\texttt{s}}$, and the reconstruction loss can be formulated as: 
\begin{equation}
    \mathcal{L}_{\texttt{rec}}=||M_{\texttt{s}}-\hat{M_{\texttt{s}}}||_{1},
\end{equation}
where $M_{\texttt{s}}$ and $\hat{M_{\texttt{s}}}$ represents the Mel-spectrogram of source waveform and predicted waveform, respectively.
To improve the performance of voice conversion, we also employ adversarial training for more natural speech. The adversarial generator loss is calculated as:
\begin{equation}
    \mathcal{L}_{\texttt{vc-adv}}^{\texttt{gen}}=(D(\hat{X_{\texttt{s}}})-1)^2,
\end{equation}
\begin{equation}
    \mathcal{L}_{\texttt{vc-adv}}^{\texttt{dis}}=(D(X_{\texttt{s}})-1)^2+D(\hat{X_{\texttt{s}}})^2,
\end{equation}
where $D(\cdot)$ is a discriminator network. For adversarial training stability, feature matching loss is also used as:
\begin{equation}
    \mathcal{L}_{\texttt{fm}}=\sum_{i=1}^T \frac{1}{N_i}\left\|D_i(X_{\texttt{s}})-D_i(\hat{X_{\texttt{s}}})\right\|_1,
\end{equation}
where T denotes the total number of layers in the discriminator and $D_i$ produces the feature map of the $i$-th layer of the discriminator with $N_i$ number of features.

Figure~\ref{fig:overview3} is an overview of the third strategy, the adversarial perturbation is directly added to the VC predicted waveform and can be defined by:
\begin{equation}
    X_{\texttt{adv}}=\hat{X}+\delta_{\texttt{adv}},
\end{equation}
where $\hat{X}$ represents the VC reconstructed waveform and $X_{\texttt{adv}}$ represents the adversarial perturbed reconstruction waveform. To adapt the input of the attacked speaker classifier, we convert the waveform to Mel-spectrogram after adding adversarial perturbation and optimize the VC model as follows:
\begin{equation}
    \mathcal{L}_{\texttt{adv}}=\left\{  
        \begin{array}{lr}  
             ||M_{\texttt{gt}}-Mel(\hat{X})||_{1}, \quad \text{if succeeded}, &  \\  
             ||Mel(X_{\texttt{adv}})-Mel(\hat{X})||_{1}, \quad \text{if failed}, &    
        \end{array}  \right.
\end{equation}
where $Mel(\cdot)$ is computing the Mel-spectrogram from waveform. When $f(Mel(\hat{X})) \neq y'$, we minimize the L1 loss between the Mel-spectrogram of the VC reconstructed waveform $Mel(\hat{X})$ and the Mel-spectrogram of the perturbed reconstruction waveform $Mel(X_{\texttt{adv}})$. For the same purpose as the former strategies, we expect the adversarial constraint can force the VC model to generate the waveform with distinctive speaker attributes.

\subsection{Generation of timbre-reserved fake audio}

\begin{figure}[ht]
  \centering
  \includegraphics[width=8.5cm]{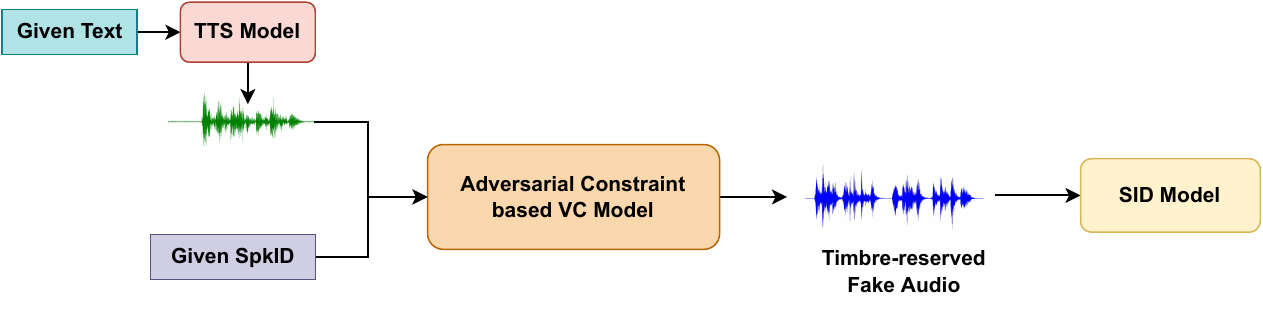}
  \caption{The flowchart of the timbre-reserved fake audio generation. }
  \label{fig:genflowchart}
\end{figure}

After the VC models with adversarial constraints are trained, the generation process of fake audio is shown in Figure~\ref{fig:genflowchart}. With the given text, we generate the audio of a random speaker by a TTS system. The TTS system is based on the FastSpeech~\cite{fs2} model and modified the encoder and decoder structure inspired by the DelightfulTTS 2~\cite{delightfultts2} conformer block. Then the given speakerID and the TTS generated audio are the inputs of the adversely constrained VC model to predict the attack Mel-spectrogram. A HifiGAN vocoder~\cite{kong2020hifi} is followed to reconstruct the waveform from Mel-spectrogram. After that, the timbre-reserved fake audios are obtained for the adversarial attack against the SID model.

\section{Experimental Setup}
\subsection{Datasets}
In this study, we use AISHELL-3~\cite{shi2020aishell} to train the speaker identification model and voice conversion model.
AISHELL-3 is a multi-speaker Mandarin Chinese audio corpus containing 88035 recordings from 218 native speakers. 
And the test set of AISHELL-1~\cite{bu2017aishell} is used to evaluate the performance of the SID model.
For the timbre-reserved adversarial attack, the dataset of audio deepfake detection (ADD) challenge~\cite{yi2022add} is used to evaluate our proposed method. 
Since ADD challenge corpus is an open-source dataset, and is specifically designed for deep fake audio attack and detection, we employ this corpus to establish a solution to first address the adversarial attack while remaining the timbre and also customizing the text. 
The training and adaptation sets for ADD Track 1 and 2 are used to train the fake detection models and the test set of these two tracks are used to evaluate the detection models.
In the test set of ADD challenge Track 3.1, 10 speaker IDs and 500 texts are given to generate the fake audio. We generate fake audio according to the given text and speaker identities and the fake audio can fool the fake detection model and SID model. Furthermore, the generated attack samples need also to meet the certain requirement of intelligibility and similarity.

\subsection{Setup}
The detailed experimental setup of all the models shown in Figure~\ref{fig:genflowchart} is described as follows:
\begin{itemize}
\item \textbf{TTS model}: The TTS model we use to generate waveforms from the given text is a 6-layer conformer encoder-decoder structure, which is similar to the DelightfulTTS 2 proposed in~\cite{delightfultts2}. With the given text, we generate the waveforms by using a randomly selected speaker timbre, and the waveforms are used as the source waveform of the VC system. 

\item \textbf{VC models}: For VC model training, a pre-trained ASR model by the WeNet toolkit~\cite{yao2021wenet} is first used to extract speaker-independent linguistic information from the source waveform and the ASR model is available on the official website~\footnote{\url{https://github.com/wenet-e2e/wenet}}. After that, the non-autoregressive and end-to-end VC models are adopted to generate the fake audios:
Firstly, the non-autoregressive VC model used to convert source speaker timbre to attack speaker timbre is an 8-layer transformer encoder-decoder structure similar to FastSpeech2~\cite{fs2} and we choose HifiGAN with multi-band processing as the vocoder~\cite{kong2020hifi}. 
On the other hand, the end-to-end VC model consists of a convolutional long short-term memory (CLSTM) encoder and a HiFiGAN-based decoder. CLSTM consists of three stacks of convolution layers followed by the LeakyReLU activation function and an LSTM layer.

After these VC models are trained, we add the speaker classifier in VC models to adversely constrain the distinctive speaker information to the generation. In the experiments of adversarial constraint on Mel-spectrogram, the learning rate $lr$ in Equation~\ref{lr} is set to be $8e-4$ and the $\delta$ is updated $1000$ times for each mini-batch. We use the $l_{\infty}$ norm to measure the perturbation bound. The $\epsilon$ starts from $0.8$.
While in the experiments on latent representation, the learning rate $lr$ is set to be $1e-4$ and $\epsilon$ starts from $0.1$. Moreover, the end-to-end VC model is trained with a $lr$ of $5e-4$ and a $\epsilon$ starting from $0.5$, and the $\delta$ is updated $2000$ times for each mini-batch. 

\item \textbf{Speaker identification model}: The SID model we used in this study is ECAPA-TDNN~\cite{desplanques2020ecapa}, which is also used as the speaker classifier in voice conversion. The EER of this model on the AISHELL-1 test set is 1.91\%. 
The architecture of ECAPA-TDNN in this study incorporates 3 SE-Res2Block modules, and channel size and the dimension of the bottleneck in the SEBlocks are set to 1024 and 256, respectively. The loss function is additive angular margin softmax (AAM-softmax)~\cite{deng2019arcface} with a margin of 0.2 and a scale of 30. 
\end{itemize}

\subsection{Evaluation metrics}
We adopt several criteria listed below to measure the performance of various kinds of generated fake audio.

\begin{itemize}

\item \textbf{Attack success rate}: is used to evaluate the performance of targeted attacks in speaker identification. The attack success rate is also the accuracy predicted from the SID, denoted as $Acc$. Formally, the accuracy is calculated as:
\begin{equation} 
    Acc = \frac{N_s}{N},
\end{equation}
where $N$ is the total number of fake audios we generated to test and $N_s$ refers to the number of audios attacking successfully.
The higher the $Acc$ in a targeted attack means the better the targeted attack is conducted.

\item \textbf{Deception success rate (DSR)}: reflects the degree of fooling the audio deepfake detection model by the generated utterances, which is defined as followed:
\begin{equation} 
    DSR = \frac{W}{A*M},
\end{equation}
where $W$ is the count of wrong detection samples by all the detection models, $A$ is the count of all the evaluation samples, and $M$ is the number of detection models.
Since we follow the rule of ADD challenge track 3.1 to generate fake audio, we also evaluate the DSR in our experiments. In order to prove that our generated timbre-reserved fake audios also have the ability to cheat the detection systems and have comparable results with the ADD challenge participating teams. 
We evaluate the fake audio on open-source deep fake detection models provided by the ASVspoof Challenge~\cite{delgado2021asvspoof} and ADD Challenge~\cite{yi2022add}, which are available on the websites: \url{https://github.com/asvspoof-challenge/2021/tree/main/DF/Baseline-RawNet2}~\cite{tak2021end} and \url{https://github.com/clovaai/aasist}~\cite{jung2022aasist}.
We further implement two similar fake detection models as the baseline systems listed in~\cite{yi2022add}, which are based on GMM and ECAPA-TDNN models. We test the EER of these two models on the test set of Track 1 and 2, which are 25.6\%, 33.1\%, 45.7\%, and 40.7\%, respectively. These are comparable to the EER of the baseline system in~\cite{yi2022add}.

\item \textbf{Objective evaluation}: 
The MOSNet~\cite{lo2019mosnet} was proposed to automatically predict the mean opinion score (MOS) of an utterance. We use the objective mean opinion score (o-MOS) predicted by the MOSNet as an objective metric of the generated fake audio quality. 
A pretrained MOSNet~\footnote{\url{https://github.com/lochenchou/MOSNet}} is used to make the o-MOS prediction, and the higher o-MOS represents the better quality of the audio.

\item \textbf{Subjective evaluation}: 
We also conduct subjective evaluations to value the fake audio from the human perceptibility of audio. The evaluation set contains 20 samples of each VC system and a total of 100 samples. 
We use comparative mean opinion score (CMOS) to compare the quality of the fake audios generated by the vanilla VC model and other adversarial based methods. 
For each pair of utterances in CMOS test, 25 native participants are asked to give a score ranging from -3 (the proposed system is much worse than baseline) to 3 (the proposed system is much better than baseline) with intervals of 1. 
Moreover, to further effectively evaluate the quality and similarity of the fake audio, we also conduct a MOS test on all the fake audio. 
Participants are asked to listen to the provided audio and evaluate their quality and similarity on a 5-point scale: 1-bad, 2-poor, 3-fair, 4-good, 5-excellent.

\item \textbf{Character error rate (CER)}: 
In order to make sure the generated fake audio meets the specific text requirements, the CER is used to evaluate the intelligibility of the fake audio, the CER is used to evaluate the intelligibility of the fake audio, and can be calculated as follows:
\begin{equation} 
    CER = \frac{S + D + I}{N},
\end{equation}
where $S$ represents the number of substitution errors, $D$ represents the number of deletion errors, $I$ is the number of insertion errors, and $N$ represents the total number of characters in the reference answer. he model we used to calculate CER is a pre-trained ASR model by the WeNet toolkit~\cite{yao2021wenet}, which is available on the official website: \url{https://github.com/wenet-e2e/wenet/tree/main/examples/wenetspeech}.

\end{itemize}

\section{Experimental Results and Analysis}
In this section, we conduct experiments to test our proposed strategies on ADD challenge dataset and then give further analyses on our method.
\subsection{Results on different reconstructed representations}

Table~\ref{tab:Accdsr} shows the attack success rates and deception success rates of the fake audios generated by various kinds of generation methods.
The first method is only generated by the vanilla VC model, which is adopted as the lower limit of all the comparisons and is denoted as `VC' in the Table. The second fake audio generation method is to add adversarial perturbation directly based on the fake audios generated by the VC model using the approach we previously proposed in~\cite{wang2020inaudible}. Since the perturbation is directly optimized by the SID system and straightforwardly added to the waveform, this method is the upper limit of the adversarial attack and is denoted as `VC+adv'. The rest three proposed methods are the VC model trained with multiple kinds of adversarial constraints as we described in Section~\ref{sec:method}, and we denote these three strategies based on the location of adversarial constraints added as `Mel', `Latent', and `Waveform', respectively. 

\subsubsection{Attack success rate}

As shown in Table~\ref{tab:Accdsr}, the attack success rate of VC model based fake audio is 29.60\%, while the VC audio with adversarial perturbation achieves 76.50\%. Meanwhile, the Acc results of our proposed strategies are 60.58\%, 54.94\%, and 66.30\%, respectively. 
We can observe that the fake audios generated based on all these three proposed timbre-reserved adversarial strategies are significantly improved compared to the fake audios generated by the original vanilla VC model, which are 30.98\%, 25.34\%, and 36.7\% absolute improvement. And our proposed methods also have comparable results with the direct perturbations to the vanilla VC generated fake audios.
Since the adversarial constraint for latent representation is difficult to have an effect on the SID, in other words, it is hard to optimize, the result of the second strategy is the lowest of these three strategies.

Due to the adversarial constraint being added only during the model training instead of directly added to the   attack waveform, the attack success rates of the proposed strategies have a small gap with the VC+adv method.
The results also illustrate that the performance varies when the adversarial constraint conducts on the different kinds of reconstructed representation. Moreover, as the reconstructed representation with adversarial constraint gets closer to the SID model, the better attack effect.

\begin{table}[h]\centering
\caption{Attack success rates (\%) and deception success rates (\%) of different kinds of generation methods.}
\label{tab:Accdsr}
\renewcommand{\tabcolsep}{0.5cm}
\renewcommand\arraystretch{1.4}
\begin{tabular}{cccc}
\bottomrule
                                              & Method     & Acc (\%) $\uparrow$      & DSR (\%) $\uparrow$      \\ \hline
Baseline                                      & VC         & 29.60          & 84.58          \\
Upper limit                                   & VC+adv     & 76.50          & 87.34          \\
\hline
\multicolumn{1}{c}{\multirow{3}{*}{\textbf{Proposed}}} & Mel & \textbf{60.58} & \textbf{90.44} \\
\multicolumn{1}{c}{}                          & Latent & \textbf{54.94} & \textbf{87.88} \\
\multicolumn{1}{c}{}                          & Waveform & \textbf{66.30} & \textbf{88.26}      \\ \bottomrule
\end{tabular}
\end{table}

\subsubsection{Deception success rate}
As shown in Table~\ref{tab:Accdsr}, the DSR results of different kinds of generation methods. We can observe that the DSR results of all these three proposed strategies outperform the vanilla VC and VC+adv method. In addition, the DSR results of all these fake audio generation methods have a comparable performance with the ADD challenge participating teams~\cite{yi2022add}, which indicates that the fake audio generated by our proposed strategies also has the ability to fool the detection systems.

\subsection{Objective and subjective evaluation}
\subsubsection{Objective evaluation}
\begin{figure}[h]
  \centering
  \includegraphics[width=8.8cm]{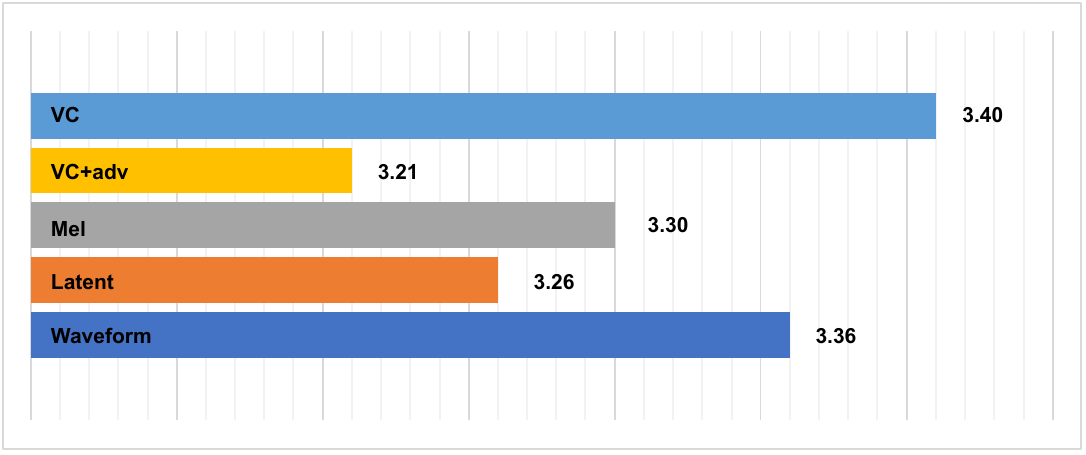}
  \caption{The o-MOS of different kinds of generation methods. }
  \label{fig:obj}
\end{figure}

The MOSNet~\cite{lo2019mosnet} prediction o-MOS is employed as the objective measurement and Figure~\ref{fig:obj} shows the objective performance of the o-MOS of various generation methods. We can observe that all the o-MOS results of fake audio generated by our proposed strategies are higher than the methods of directly adding adversarial perturbation to the audio. This indicates that compared to methods of adding adversarial perturbation directly to the audio, the fake audios generated based on our proposed strategies have better quality since we add adversarial constraints to the VC model avoiding directly introducing extra perturbations to the fake audio.

\subsubsection{Subjective evaluation}

We conduct a CMOS evaluation shown in Table~\ref{tab:obj}. The result shows that our proposed strategies achieve 0.22, 0.20, and 0.17 CMOS higher than the VC+adv method, demonstrating the fake audio generated by our proposed strategies is more similar to audio generated by the vanilla VC model. Compared to the VC+adv method introducing extra noise to the waveform, our proposed strategies can get rid of the influence of the additional perturbation by adding adversarial constraints in the generation model.

We also conduct MOS tests on various generation methods, to evaluate the quality and speaker similarity of all kinds of fake audio. The results of the MOS test are shown in Table~\ref{tab:obj}. 
We can observe from the MOS results of quality that our proposed strategies outperform the VC+adv method, indicating that the audio quality generated from our proposed strategies is better than the VC+adv method, while is comparable to the quality of the audio generated by the vanilla VC model. This demonstrates that the adversarial constraint added during the VC model training does not influence the quality of the generated audio.
For the speaker similarity, the MOS results of our proposed strategies are also slightly higher than the VC+adv method.

These subjective results demonstrate the proposed strategies based on adversarial constraints can avoid the quality damage caused by extra adversarial perturbations and also do not influence the capability of the VC model.

\begin{table}[h]\centering
\caption{CMOS and MOS results of fake audios generated by different strategies.}
\label{tab:obj}
\renewcommand{\tabcolsep}{0.5cm}
\renewcommand\arraystretch{1.3}
\begin{tabular}{cccc}
\bottomrule
\multirow{2}{*}{} & \multirow{2}{*}{CMOS $\uparrow$} & \multicolumn{2}{c}{MOS $\uparrow$} \\ \cline{3-4} 
                  &                       & Quality   & Similarity  \\ \hline
VC                & -                     & 3.91$\pm0.07$      & 3.87$\pm0.06$        \\
VC+adv            & -0.34                 & 3.42$\pm0.11$      & 3.79$\pm0.08$        \\
\hline
\textbf{Mel}           & -0.12                 & 3.85$\pm0.08$      & 3.86$\pm0.06$        \\
\textbf{Latent}           & -0.14                 & 3.82$\pm0.08$      & 3.84$\pm0.07$        \\
\textbf{Waveform}           & -0.17                 & 3.78$\pm0.09$      & 3.88$\pm0.07$        \\ \bottomrule
\end{tabular}
\end{table}

\subsection{Analysis}
\subsubsection{Speaker similarity of fake audios}
\begin{figure}[h]
  \centering
  \includegraphics[width=8.86cm]{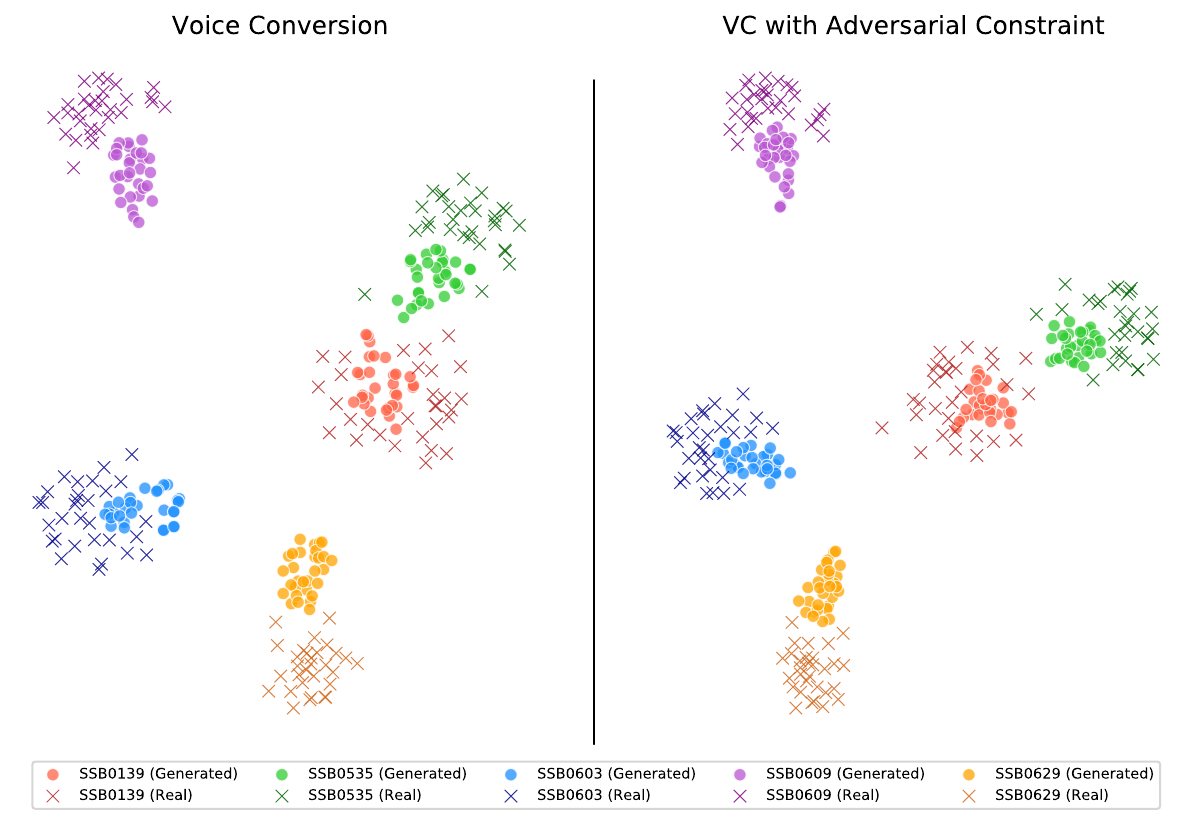}
  \caption{Visualization of male speaker distributions. }
  \label{fig:spkvisual}
\end{figure}

\begin{figure}[h]
  \centering
  \includegraphics[width=8.86cm]{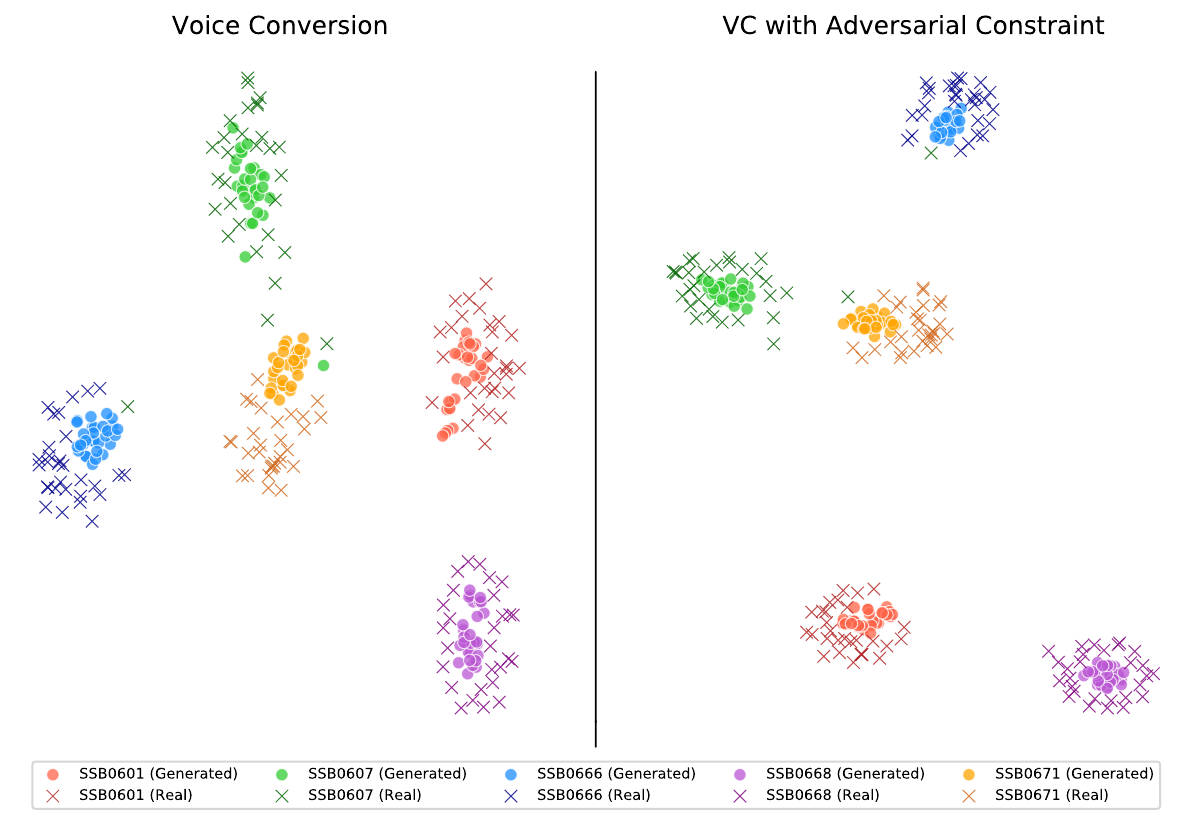}
  \caption{Visualization of female speaker distributions. }
  \label{fig:spkvisual2}
\end{figure}

As shown in Figure~\ref{fig:spkvisual} and~\ref{fig:spkvisual2}, we utilize a pre-trained  speaker encoder~\cite{desplanques2020ecapa} to extract speaker embeddings of real and fake speech, which are visualized through t-SNE~\cite{van2008visualizing}.
The distribution of male and female speakers are respectively visualized. In these two figures, all the `x' represents the real data from various speakers, while all the `o' represents the generated fake audio. Different colors on behalf of different speakers. In particular, the fake audios in the left parts of both figures are generated by the vanilla non-autoregressive based VC model, while those in the right are generated by the VC model with adversarial constraint.

We can observe the fake audio samples generated by the VC with adversarial constraints are more clustered and nearer to the real audio samples than the samples generated by the original VC model. It means the similarity of the fake audio generated by the proposed method is closer to the real audio of target speakers, since the adversarial constraints exploit the weakness of SID model.

From the analysis above, the fake audios generated by our proposed methods are not only timbre-reserved but also more speaker-wise than the original vanilla VC model. The speaker similarity is improved after the adversarial constraint, which is beneficial to successful attacks.

\subsubsection{Intelligibility of fake audios}
The CER results are shown in Table~\ref{tab:CER}. Since our proposed adversarial attack is also text customized, we can see that the CERs of all kinds of fake audio generation methods are low, which indicates that the text of the fake audio can meet specific requirements. In the specified attack scenarios, we also can customize the text. 

\begin{table}[h]\centering
\caption{CERs (\%) of fake audios generated by different methods.}
\label{tab:CER}
\begin{tabular}{c|ccccl}
\bottomrule
Method & VC    & VC+adv & Mel & Latent & Waveform \\ \hline
CER (\%) $\downarrow$    & 3.97 & 3.57  & 3.93      & 3.97      &  3.99      \\ 
\bottomrule
\end{tabular}
\end{table}

From the discussion mentioned above, both speaker similarity and intelligibility are indicated to be guaranteed.
\subsection{Comparison of different kinds of constraints}
Cai~\textit{et al.}~\cite{cai2020speaker} added a speaker identity-related loss to constrain the centralized to improve the speaker similarity between the synthesized speech and its natural reference audio. This feedback constraint (FC) also can be added to VC model to constrain the distance of the VC predicted Mel-spectrogram and the real Mel-spectrogram. 
As shown in Table~\ref{tab:cai}, we can observe that our proposed adversarial constraint based on Mel-spectrogram outperforms the feedback constraint (FC). Since the adversarial constraints exploit the weakness of SID model, it is more beneficial to adversarial attack, while the FC mainly focuses on generating Mel-spectrogram fitting the real Mel-spectrogram as much as possible.
\begin{table}[h]\centering
\caption{Comparison of adversarial constraint and feedback constraint~\cite{cai2020speaker}.}
\label{tab:cai}
\renewcommand{\tabcolsep}{0.5cm}
\renewcommand\arraystretch{1.4}
\begin{tabular}{ccc}
\bottomrule
                                              Method     & Acc (\%) $\uparrow$      & DSR (\%) $\uparrow$      \\ \hline
Adversarial constraint (Mel) & 60.58 & 90.44 \\
Feedback constraint~\cite{cai2020speaker} & 47.84 & 88.32\\ \bottomrule
\end{tabular}
\end{table}
\section{Conclusion}
In this study, we propose to generate timbre-reserved fake audios for the adversarial attack in the speaker identification system. By adding adversarial constraints to different levels of representations of the VC model, we can generate timbre-reserved and speaker-wised fake audio to attack the SID model.
Experiments on ADD challenge corpus show that our proposed strategies significantly improve the attack success rate compared to the vanilla VC model.
The objective and subjective evaluation demonstrate that adversarial constraints do not affect the quality of the VC model and also can get rid of the influence of the extra noise by adversely constraining the fake audio generation model instead of directly adding adversarial perturbation to the waveform.
Moreover, we also analyze the speaker similarity and intelligibility of the fake audio, the speaker similarity is improved after the adversarial constraint, which is beneficial to attack successfully. And this can also prove that our generated timbre-reserved fake audio is speaker-wised and the text of our adversarial attack can be customized.

%
\IEEEpeerreviewmaketitle


%

\appendices




\ifCLASSOPTIONcaptionsoff
  \newpage
\fi



\bibliographystyle{IEEEtran}
%
\bibliography{mybib}

\begin{thebibliography}{10}
\providecommand{\url}[1]{#1}
\csname url@samestyle\endcsname
\providecommand{\newblock}{\relax}
\providecommand{\bibinfo}[2]{#2}
\providecommand{\BIBentrySTDinterwordspacing}{\spaceskip=0pt\relax}
\providecommand{\BIBentryALTinterwordstretchfactor}{4}
\providecommand{\BIBentryALTinterwordspacing}{\spaceskip=\fontdimen2\font plus
\BIBentryALTinterwordstretchfactor\fontdimen3\font minus
  \fontdimen4\font\relax}
\providecommand{\BIBforeignlanguage}[2]{{%
\expandafter\ifx\csname l@#1\endcsname\relax
\typeout{** WARNING: IEEEtran.bst: No hyphenation pattern has been}%
\typeout{** loaded for the language `#1'. Using the pattern for}%
\typeout{** the default language instead.}%
\else
\language=\csname l@#1\endcsname
\fi
#2}}
\providecommand{\BIBdecl}{\relax}
\BIBdecl

\bibitem{hansen2015speaker}
J.~H. Hansen and T.~Hasan, ``Speaker recognition by machines and humans: A
  tutorial review,'' \emph{IEEE Signal Processing Magazine}, vol.~32, no.~6,
  pp. 74--99, 2015.

\bibitem{reynolds1995robust}
D.~A. Reynolds and R.~C. Rose, ``Robust text-independent speaker identification
  using gaussian mixture speaker models,'' \emph{IEEE Transactions on Speech
  and Audio Processing}, vol.~3, no.~1, pp. 72--83, 1995.

\bibitem{wu2012detecting}
Z.~Wu, E.~S. Chng, and H.~Li, ``Detecting converted speech and natural speech
  for anti-spoofing attack in speaker recognition,'' in \emph{Proc.
  INTERSPEECH}, 2012.

\bibitem{wu2015spoofing}
Z.~Wu, N.~Evans, T.~Kinnunen, J.~Yamagishi, F.~Alegre, and H.~Li, ``Spoofing
  and countermeasures for speaker verification: a survey,'' \emph{Speech
  Communication}, vol.~66, pp. 130--153, 2015.

\bibitem{wu2017asvspoof}
Z.~Wu, J.~Yamagishi, T.~Kinnunen, C.~Hanil{\c{c}}i, M.~Sahidullah, A.~Sizov,
  N.~Evans, M.~Todisco, and H.~Delgado, ``A{SV}spoof: the automatic speaker
  verification spoofing and countermeasures challenge,'' \emph{IEEE Journal of
  Selected Topics in Signal Processing}, vol.~11, no.~4, pp. 588--604, 2017.

\bibitem{wu2015asvspoof}
Z.~Wu, T.~Kinnunen, N.~Evans, J.~Yamagishi, C.~Hanil{\c{c}}i, M.~Sahidullah,
  and A.~Sizov, ``A{SV}spoof 2015: the first automatic speaker verification
  spoofing and countermeasures challenge,'' in \emph{Proc. INTERSPEECH}, 2015.

\bibitem{kinnunen2017asvspoof}
T.~Kinnunen, M.~Sahidullah, H.~Delgado, M.~Todisco, N.~Evans, J.~Yamagishi, and
  K.~A. Lee, ``The {ASV}spoof 2017 challenge: Assessing the limits of replay
  spoofing attack detection,'' 2017.

\bibitem{todisco2019asvspoof}
M.~Todisco, X.~Wang, V.~Vestman, M.~Sahidullah, H.~Delgado, A.~Nautsch,
  J.~Yamagishi, N.~Evans, T.~Kinnunen, and K.~A. Lee, ``A{SV}spoof 2019: Future
  horizons in spoofed and fake audio detection,'' \emph{arXiv preprint
  arXiv:1904.05441}, 2019.

\bibitem{goodfellow2014explaining}
I.~J. Goodfellow, J.~Shlens, and C.~Szegedy, ``Explaining and harnessing
  adversarial examples,'' \emph{arXiv preprint arXiv:1412.6572}, 2014.

\bibitem{carlini2017towards}
N.~Carlini and D.~Wagner, ``Towards evaluating the robustness of neural
  networks,'' in \emph{Proc. SP)}.\hskip 1em plus 0.5em minus 0.4em\relax IEEE,
  2017, pp. 39--57.

\bibitem{szegedy2013intriguing}
C.~Szegedy, W.~Zaremba, I.~Sutskever, J.~Bruna, D.~Erhan, I.~Goodfellow, and
  R.~Fergus, ``Intriguing properties of neural networks,'' \emph{arXiv preprint
  arXiv:1312.6199}, 2013.

\bibitem{kurakin2016adversarial}
A.~Kurakin, I.~Goodfellow, and S.~Bengio, ``Adversarial examples in the
  physical world,'' \emph{arXiv preprint arXiv:1607.02533}, 2016.

\bibitem{carlini2018audio}
N.~Carlini and D.~Wagner, ``Audio adversarial examples: Targeted attacks on
  speech-to-text,'' in \emph{2018 IEEE Security and Privacy Workshops (SPW)},
  2018, pp. 1--7.

\bibitem{schonherr2018adversarial}
L.~Sch{\"o}nherr, K.~Kohls, S.~Zeiler, T.~Holz, and D.~Kolossa, ``Adversarial
  attacks against automatic speech recognition systems via psychoacoustic
  hiding,'' in \emph{Proc. NDSS}, 2019.

\bibitem{alzantot2018did}
M.~Alzantot, B.~Balaji, and M.~Srivastava, ``Did you hear that? {a}dversarial
  examples against automatic speech recognition,'' \emph{arXiv preprint
  arXiv:1801.00554}, 2018.

\bibitem{sun2019adversarial}
S.~Sun, P.~Guo, L.~Xie, and M.-Y. Hwang, ``Adversarial regularization for
  attention based end-to-end robust speech recognition,'' \emph{IEEE/ACM
  Transactions on Audio, Speech and Language Processing (TASLP)}, vol.~27,
  no.~11, pp. 1826--1838, 2019.

\bibitem{qin2019imperceptible}
Y.~Qin, N.~Carlini, I.~Goodfellow, G.~Cottrell, and C.~Raffel, ``Imperceptible,
  robust, and targeted adversarial examples for automatic speech recognition,''
  in \emph{Proc. ICML)}.\hskip 1em plus 0.5em minus 0.4em\relax PMLR, 2019, pp.
  5231--5240.

\bibitem{abdullah2021sok}
H.~Abdullah, K.~Warren, V.~Bindschaedler, N.~Papernot, and P.~Traynor, ``So{K}:
  The faults in our {ASR}s: An overview of attacks against automatic speech
  recognition and speaker identification systems,'' in \emph{Proc. SP}.\hskip
  1em plus 0.5em minus 0.4em\relax IEEE, 2021, pp. 730--747.

\bibitem{snyder2018x}
D.~Snyder, D.~Garcia-Romero, G.~Sell, D.~Povey, and S.~Khudanpur, ``X-vectors:
  Robust {DNN} embeddings for speaker recognition,'' in \emph{Proc.
  ICASSP}.\hskip 1em plus 0.5em minus 0.4em\relax IEEE, 2018, pp. 5329--5333.

\bibitem{wan2018generalized}
L.~Wan, Q.~Wang, A.~Papir, and I.~L. Moreno, ``Generalized end-to-end loss for
  speaker verification,'' in \emph{Proc. ICASSP}.\hskip 1em plus 0.5em minus
  0.4em\relax IEEE, 2018, pp. 4879--4883.

\bibitem{desplanques2020ecapa}
B.~Desplanques, J.~Thienpondt, and K.~Demuynck, ``{ECAPA-TDNN}: Emphasized
  channel attention, propagation and aggregation in {TDNN} based speaker
  verification,'' \emph{arXiv preprint arXiv:2005.07143}, 2020.

\bibitem{kreuk2018fooling}
F.~Kreuk, Y.~Adi, M.~Cisse, and J.~Keshet, ``Fooling end-to-end speaker
  verification with adversarial examples,'' in \emph{Proc. ICASSP}.\hskip 1em
  plus 0.5em minus 0.4em\relax IEEE, 2018, pp. 1962--1966.

\bibitem{wang2019adversarial}
Q.~Wang, P.~Guo, S.~Sun, L.~Xie, and J.~H. Hansen, ``Adversarial regularization
  for end-to-end robust speaker verification,'' in \emph{Proc. INTERSPEECH},
  2019, pp. 4010--4014.

\bibitem{abdullah2019practical}
H.~Abdullah, W.~Garcia, C.~Peeters, P.~Traynor, K.~R. Butler, and J.~Wilson,
  ``Practical hidden voice attacks against speech and speaker recognition
  systems,'' \emph{arXiv preprint arXiv:1904.05734}, 2019.

\bibitem{li2020universal}
J.~Li, X.~Zhang, C.~Jia, J.~Xu, L.~Zhang, Y.~Wang, S.~Ma, and W.~Gao,
  ``Universal adversarial perturbations generative network for speaker
  recognition,'' in \emph{Proc. ICME}.\hskip 1em plus 0.5em minus 0.4em\relax
  IEEE, 2020, pp. 1--6.

\bibitem{xie2020real}
Y.~Xie, C.~Shi, Z.~Li, J.~Liu, Y.~Chen, and B.~Yuan, ``Real-time, universal,
  and robust adversarial attacks against speaker recognition systems,'' in
  \emph{Proc. ICASSP}.\hskip 1em plus 0.5em minus 0.4em\relax IEEE, 2020, pp.
  1738--1742.

\bibitem{li2020practical}
Z.~Li, C.~Shi, Y.~Xie, J.~Liu, B.~Yuan, and Y.~Chen, ``Practical adversarial
  attacks against speaker recognition systems,'' in \emph{Proc. HOTMOBILE},
  2020, pp. 9--14.

\bibitem{li2020adversarial}
X.~Li, J.~Zhong, X.~Wu, J.~Yu, X.~Liu, and H.~Meng, ``Adversarial attacks on
  {GMM} i-vector based speaker verification systems,'' in \emph{Proc.
  ICASSP}.\hskip 1em plus 0.5em minus 0.4em\relax IEEE, 2020, pp. 6579--6583.

\bibitem{chen2021real}
G.~Chen, S.~Chenb, L.~Fan, X.~Du, Z.~Zhao, F.~Song, and Y.~Liu, ``Who is real
  bob? adversarial attacks on speaker recognition systems,'' in \emph{Proc.
  SP}.\hskip 1em plus 0.5em minus 0.4em\relax IEEE, 2021, pp. 694--711.

\bibitem{jati2021adversarial}
A.~Jati, C.-C. Hsu, M.~Pal, R.~Peri, W.~AbdAlmageed, and S.~Narayanan,
  ``Adversarial attack and defense strategies for deep speaker recognition
  systems,'' \emph{Computer Speech \& Language}, vol.~68, p. 101199, 2021.

\bibitem{wang2020inaudible}
Q.~Wang, P.~Guo, and L.~Xie, ``Inaudible adversarial perturbations for targeted
  attack in speaker recognition,'' \emph{Proc. INTERSPEECH}, pp. 4228--4232,
  2020.

\bibitem{zhang2022imperceptible}
X.~Zhang, X.~Zhang, M.~Sun, X.~Zou, K.~Chen, and N.~Yu, ``Imperceptible
  black-box waveform-level adversarial attack towards automatic speaker
  recognition,'' \emph{Complex \& Intelligent Systems}, pp. 1--15, 2022.

\bibitem{yi2022add}
J.~Yi, R.~Fu, J.~Tao, S.~Nie, H.~Ma, C.~Wang, T.~Wang, Z.~Tian, Y.~Bai, C.~Fan
  \emph{et~al.}, ``{ADD} 2022: the first audio deep synthesis detection
  challenge,'' in \emph{Proc. ICASSP}.\hskip 1em plus 0.5em minus 0.4em\relax
  IEEE, 2022, pp. 9216--9220.

\bibitem{wang2017tacotron}
Y.~Wang, R.~Skerry-Ryan, D.~Stanton, Y.~Wu, R.~J. Weiss, N.~Jaitly, Z.~Yang,
  Y.~Xiao, Z.~Chen, S.~Bengio \emph{et~al.}, ``Tacotron: Towards end-to-end
  speech synthesis,'' \emph{arXiv preprint arXiv:1703.10135}, 2017.

\bibitem{ren2019fastspeech}
Y.~Ren, Y.~Ruan, X.~Tan, T.~Qin, S.~Zhao, Z.~Zhao, and T.-Y. Liu, ``Fastspeech:
  Fast, robust and controllable text to speech,'' \emph{Proc. NeurIPS},
  vol.~32, 2019.

\bibitem{chen2010speaker}
L.-W. Chen, W.~Guo, and L.-R. Dai, ``Speaker verification against synthetic
  speech,'' in \emph{Proc. ISCSLP}.\hskip 1em plus 0.5em minus 0.4em\relax
  IEEE, 2010, pp. 309--312.

\bibitem{shchemelinin2014vulnerability}
V.~Shchemelinin, M.~Topchina, and K.~Simonchik, ``Vulnerability of voice
  verification systems to spoofing attacks by {TTS} voices based on
  automatically labeled telephone speech,'' in \emph{Proc. SPECOM}.\hskip 1em
  plus 0.5em minus 0.4em\relax Springer, 2014, pp. 475--481.

\bibitem{cai2018attacking}
W.~Cai, A.~Doshi, and R.~Valle, ``Attacking speaker recognition with deep
  generative models,'' \emph{arXiv preprint arXiv:1801.02384}, 2018.

\bibitem{mehri2016samplernn}
S.~Mehri, K.~Kumar, I.~Gulrajani, R.~Kumar, S.~Jain, J.~Sotelo, A.~Courville,
  and Y.~Bengio, ``Sample{RNN}: An unconditional end-to-end neural audio
  generation model,'' \emph{arXiv preprint arXiv:1612.07837}, 2016.

\bibitem{oord2016wavenet}
A.~v.~d. Oord, S.~Dieleman, H.~Zen, K.~Simonyan, O.~Vinyals, A.~Graves,
  N.~Kalchbrenner, A.~Senior, and K.~Kavukcuoglu, ``Wavenet: A generative model
  for raw audio,'' \emph{arXiv preprint arXiv:1609.03499}, 2016.

\bibitem{wenger2021hello}
E.~Wenger, M.~Bronckers, C.~Cianfarani, J.~Cryan, A.~Sha, H.~Zheng, and B.~Y.
  Zhao, ```` {H}ello, {I}t's {M}e": Deep learning-based speech synthesis
  attacks in the real world,'' in \emph{Proc. CCS}.\hskip 1em plus 0.5em minus
  0.4em\relax ACM, 2021, pp. 235--251.

\bibitem{sisman2020overview}
B.~Sisman, J.~Yamagishi, S.~King, and H.~Li, ``An overview of voice conversion
  and its challenges: From statistical modeling to deep learning,''
  \emph{IEEE/ACM Transactions on Audio, Speech, and Language Processing
  (TASLP)}, vol.~29, pp. 132--157, 2020.

\bibitem{sun2015voice}
L.~Sun, S.~Kang, K.~Li, and H.~Meng, ``Voice conversion using deep
  bidirectional long short-term memory based recurrent neural networks,'' in
  \emph{Proc. ICASSP}.\hskip 1em plus 0.5em minus 0.4em\relax IEEE, 2015, pp.
  4869--4873.

\bibitem{sun2016phonetic}
L.~Sun, K.~Li, H.~Wang, S.~Kang, and H.~Meng, ``Phonetic posteriorgrams for
  many-to-one voice conversion without parallel data training,'' in
  \emph{Proc.ICME}.\hskip 1em plus 0.5em minus 0.4em\relax IEEE, 2016, pp.
  1--6.

\bibitem{hayashi2021non}
T.~Hayashi, W.-C. Huang, K.~Kobayashi, and T.~Toda, ``Non-autoregressive
  sequence-to-sequence voice conversion,'' in \emph{Proc. ICASSP}.\hskip 1em
  plus 0.5em minus 0.4em\relax IEEE, 2021, pp. 7068--7072.

\bibitem{liu2020non}
L.-J. Liu, Y.-N. Chen, J.-X. Zhang, Y.~Jiang, Y.-J. Hu, Z.-H. Ling, and L.-R.
  Dai, ``Non-parallel voice conversion with autoregressive conversion model and
  duration adjustment,'' in \emph{Proc. Joint Workshop for the Blizzard
  Challenge and Voice Conversion Challenge 2020}, 2020, pp. 126--130.

\bibitem{hayashi2022investigation}
T.~Hayashi, K.~Kobayashi, and T.~Toda, ``An investigation of streaming
  non-autoregressive sequence-to-sequence voice conversion,'' in \emph{Proc.
  ICASSP}.\hskip 1em plus 0.5em minus 0.4em\relax IEEE, 2022, pp. 6802--6806.

\bibitem{kameoka2018stargan}
H.~Kameoka, T.~Kaneko, K.~Tanaka, and N.~Hojo, ``Stargan-vc: Non-parallel
  many-to-many voice conversion using star generative adversarial networks,''
  in \emph{Proc. SLT}.\hskip 1em plus 0.5em minus 0.4em\relax IEEE, 2018, pp.
  266--273.

\bibitem{kaneko2018cyclegan}
T.~Kaneko and H.~Kameoka, ``Cyclegan-vc: Non-parallel voice conversion using
  cycle-consistent adversarial networks,'' in \emph{Proc. EUSIPCO}.\hskip 1em
  plus 0.5em minus 0.4em\relax IEEE, 2018, pp. 2100--2104.

\bibitem{lin2021fragmentvc}
Y.~Y. Lin, C.-M. Chien, J.-H. Lin, H.-y. Lee, and L.-s. Lee, ``Fragmentvc:
  Any-to-any voice conversion by end-to-end extracting and fusing fine-grained
  voice fragments with attention,'' in \emph{Proc. ICASSP}.\hskip 1em plus
  0.5em minus 0.4em\relax IEEE, 2021, pp. 5939--5943.

\bibitem{nguyen2022nvc}
B.~Nguyen and F.~Cardinaux, ``{NVC-NET}: End-to-end adversarial voice
  conversion,'' in \emph{Proc. ICASSP}.\hskip 1em plus 0.5em minus 0.4em\relax
  IEEE, 2022, pp. 7012--7016.

\bibitem{kim2021conditional}
J.~Kim, J.~Kong, and J.~Son, ``Conditional variational autoencoder with
  adversarial learning for end-to-end text-to-speech,'' in \emph{Proc.
  ICML}.\hskip 1em plus 0.5em minus 0.4em\relax PMLR, 2021, pp. 5530--5540.

\bibitem{shen2018natural}
J.~Shen, R.~Pang, R.~J. Weiss, M.~Schuster, N.~Jaitly, Z.~Yang, Z.~Chen,
  Y.~Zhang, Y.~Wang, R.~Skerrv-Ryan \emph{et~al.}, ``Natural {TTS} synthesis by
  conditioning wavenet on {Mel} spectrogram predictions,'' in \emph{Proc.
  ICASSP}.\hskip 1em plus 0.5em minus 0.4em\relax IEEE, 2018, pp. 4779--4783.

\bibitem{fs2}
Y.~Ren, C.~Hu, X.~Tan, T.~Qin, S.~Zhao, Z.~Zhao, and T.-Y. Liu, ``Fastspeech 2:
  Fast and high-quality end-to-end text to speech,'' in \emph{Proc. ICLR},
  2021.

\bibitem{kashkin2022hifi}
A.~Kashkin, I.~Karpukhin, and S.~Shishkin, ``Hi{F}i-{VC}: High quality
  asr-based voice conversion,'' \emph{arXiv preprint arXiv:2203.16937}, 2022.

\bibitem{das2020attacker}
R.~K. Das, X.~Tian, T.~Kinnunen, and H.~Li, ``The attacker's perspective on
  automatic speaker verification: an overview,'' \emph{Proc. INTERSPEECH}, pp.
  4213--4217, 2020.

\bibitem{kong2020hifi}
J.~Kong, J.~Kim, and J.~Bae, ``Hi{F}i-{GAN}: Generative adversarial networks
  for efficient and high fidelity speech synthesis,'' \emph{Proc. NeurIPS},
  vol.~33, pp. 17\,022--17\,033, 2020.

\bibitem{delightfultts2}
Y.~Liu, R.~Xue, L.~He, X.~Tan, and S.~Zhao, ``Delightfultts 2: End-to-end
  speech synthesis with adversarial vector-quantized auto-encoders,'' in
  \emph{Proc. Interspeech}, 2022, pp. 1581--1585.

\bibitem{shi2020aishell}
Y.~Shi, H.~Bu, X.~Xu, S.~Zhang, and M.~Li, ``Aishell-3: A multi-speaker
  mandarin tts corpus and the baselines,'' \emph{arXiv preprint
  arXiv:2010.11567}, 2020.

\bibitem{bu2017aishell}
H.~Bu, J.~Du, X.~Na, B.~Wu, and H.~Zheng, ``Aishell-1: an open-source mandarin
  speech corpus and a speech recognition baseline,'' in \emph{Proc.
  O-COCOSDA}.\hskip 1em plus 0.5em minus 0.4em\relax IEEE, 2017, pp. 1--5.

\bibitem{yao2021wenet}
Z.~Yao, D.~Wu, X.~Wang, B.~Zhang, F.~Yu, C.~Yang, Z.~Peng, X.~Chen, L.~Xie, and
  X.~Lei, ``Wenet: Production oriented streaming and non-streaming end-to-end
  speech recognition toolkit,'' in \emph{Proc. Interspeech}, 2021, pp.
  4054--4058.

\bibitem{deng2019arcface}
J.~Deng, J.~Guo, N.~Xue, and S.~Zafeiriou, ``Arcface: Additive angular margin
  loss for deep face recognition,'' in \emph{Proc. CVPR}.\hskip 1em plus 0.5em
  minus 0.4em\relax IEEE/CVF, 2019, pp. 4690--4699.

\bibitem{delgado2021asvspoof}
H.~Delgado, N.~Evans, T.~Kinnunen, K.~A. Lee, X.~Liu, A.~Nautsch, J.~Patino,
  M.~Sahidullah, M.~Todisco, X.~Wang \emph{et~al.}, ``Asvspoof 2021: Automatic
  speaker verification spoofing and countermeasures challenge evaluation
  plan,'' \emph{arXiv preprint arXiv:2109.00535}, 2021.

\bibitem{tak2021end}
H.~Tak, J.-w. Jung, J.~Patino, M.~Kamble, M.~Todisco, and N.~Evans,
  ``End-to-end spectro-temporal graph attention networks for speaker
  verification anti-spoofing and speech deepfake detection,'' \emph{arXiv
  preprint arXiv:2107.12710}, 2021.

\bibitem{jung2022aasist}
J.-w. Jung, H.-S. Heo, H.~Tak, H.-j. Shim, J.~S. Chung, B.-J. Lee, H.-J. Yu,
  and N.~Evans, ``Aasist: Audio anti-spoofing using integrated spectro-temporal
  graph attention networks,'' in \emph{Proc. ICASSP}.\hskip 1em plus 0.5em
  minus 0.4em\relax IEEE, 2022, pp. 6367--6371.

\bibitem{lo2019mosnet}
C.-C. Lo, S.-W. Fu, W.-C. Huang, X.~Wang, J.~Yamagishi, Y.~Tsao, and H.-M.
  Wang, ``Mosnet: Deep learning based objective assessment for voice
  conversion,'' \emph{arXiv preprint arXiv:1904.08352}, 2019.

\bibitem{van2008visualizing}
L.~Van~der Maaten and G.~Hinton, ``Visualizing data using t-sne.''
  \emph{Journal of machine learning research}, vol.~9, no.~11, 2008.

\bibitem{cai2020speaker}
Z.~Cai, C.~Zhang, and M.~Li, ``From speaker verification to multispeaker speech
  synthesis, deep transfer with feedback constraint,'' \emph{arXiv preprint
  arXiv:2005.04587}, 2020.

\end{thebibliography}
\end{document}